\documentclass[superscriptaddress, twocolumn,aps, longbibliography]{revtex4-2}
\usepackage[utf8]{inputenc}
\usepackage[T1]{fontenc}
\usepackage{amsmath, amssymb, color, stmaryrd, wasysym, esint}
\usepackage{scalerel}
\makeatletter
\newsavebox{\@brx}
\newcommand{\llangle}[1][]{\savebox{\@brx}{\(\m@th{#1\langle}\)}%
  \mathopen{\copy\@brx\kern-0.5\wd\@brx\usebox{\@brx}}}
\newcommand{\rrangle}[1][]{\savebox{\@brx}{\(\m@th{#1\rangle}\)}%
  \mathclose{\copy\@brx\kern-0.5\wd\@brx\usebox{\@brx}}}
\makeatother

\makeatletter
\newsavebox{\@brxx}
\newcommand{\lllangle}[1][]{\savebox{\@brxx}{\(\m@th{#1\langle}\)}%
  \mathopen{\copy\@brxx\kern-0.5\wd\@brxx\usebox{\@brxx}\kern-0.5\wd\@brxx\usebox{\@brxx}}}
\newcommand{\rrrangle}[1][]{\savebox{\@brxx}{\(\m@th{#1\rangle}\)}%
  \mathclose{\copy\@brxx\kern-0.5\wd\@brxx\usebox{\@brxx}\kern-0.5\wd\@brxx\usebox{\@brxx}}}
\makeatother

\usepackage{graphicx}
\usepackage{dcolumn}
\usepackage{bm}
\usepackage{xcolor}
\usepackage{xspace}
\usepackage[makeroom]{cancel}

\usepackage{float}
\usepackage{siunitx}
\usepackage{mathtools}
\usepackage{algorithm}
\usepackage{algorithmic}

\definecolor{linkcolor}{rgb}{0,0,0.6} 
\usepackage[pdftex,colorlinks=true,
	pdfstartview = FitV,
	linkcolor    = linkcolor,
	citecolor    = linkcolor,
	urlcolor     = linkcolor,
	hyperindex   = true,
	hyperfigures = false]{hyperref}

\begin{document}
\title{Insertion space in repulsive active matter}

\author{Luke K. Davis}
\email{luke.davis@ed.ac.uk} 
\affiliation{%
School of Mathematics and Maxwell Institute for Mathematical Sciences, University of Edinburgh, EH9 3FD, Scotland
}%
\affiliation{%
Higgs Centre for Theoretical Physics, University of Edinburgh, EH9 3FD, Scotland
}%
\affiliation{%
Isaac Newton Institute for Mathematical Sciences, University of Cambridge, CB3 0EH, England
}%
\author{Karel Proesmans} 
\affiliation{%
Niels Bohr International Academy, Niels Bohr Institute, Blegdamsvej 17, DK-2100 Copenhagen
}%

\begin{abstract}
For equilibrium hard spheres the statistical geometry of the insertion space, the room to accommodate another sphere, relates exactly to the equation of state. We begin to extend this idea to active matter, analyzing the insertion space for repulsive active particles in one and two dimensions using both on- and off-lattice models. In one-dimension we derive closed-form expressions for the mean insertion cavity size, cavity number, and total insertion volume, all in excellent agreement with simulations. Strikingly, activity increases the total insertion volume and tends to keep the insertion space more connected. We also find that insertion space metrics contain signatures of collective phase behaviors occurring at previously predicted packing fractions. Taken together, our work provides the first quantitative foundation for the statistical geometry of active matter.
\end{abstract}

\maketitle

\begin{figure}[t!]
    \centering
    \includegraphics[width=0.7\linewidth]{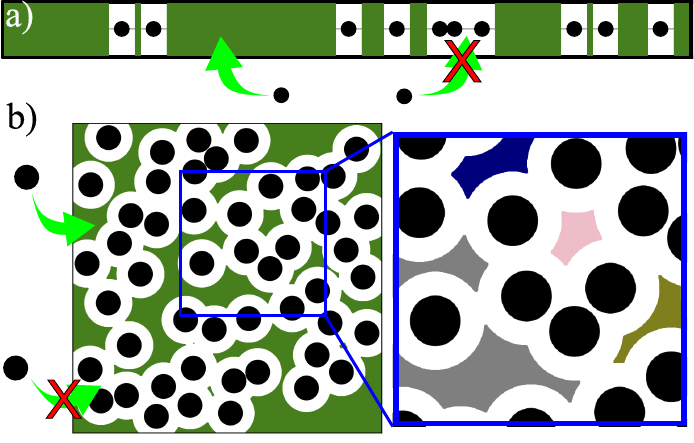}
    \caption{ a) Visualized insertion cavities for a configuration of hard spheres in $d=1$. b) Example insertion cavities for a configuration of hard discs in a closed box. Disconnected cavities identified in our algorithm (right).}
    \label{fig:ISExample}
\end{figure}

Insertion is one of the most basic actions in the natural and human worlds: it is the act of storing physical objects in space. The practical difficulty of insertion lies in the objects taking up a limited amount of space and accounting for the many ways to arrange the objects (Fig. \ref{fig:ISExample}). Since the work of B. Widom (1960s), only insertion into a medium in equilibrium with its environment has been considered \cite{Widom1963,Speedy1977}, but what happens if the insertion medium is, instead, motile and far-from-equilibrium? Understanding insertion has important practical ramifications regarding the storage of motile units, such as self-propelled colloids \cite{Bishop2023}, robots \cite{BaronaBalda2024}, and biological cells \cite{Needleman2017}, which could form the next generation of engines \cite{Pietzonka2019}. Here, we provide the first answer to this question in the context of active matter: a broad class of non-equilibrium systems whose constituents turnover energy to fuel their dynamics \cite{Marchetti2013}. 

Active constituents interact in many ways, with volume-exclusion (a form of repulsion) being the most relevant for insertion \cite{Tonks1936,ArnoulxdePirey2019,Royall2024}. An important phenomenon that arises for active--as opposed to passive--repulsive particles is motility-induced phase separation (MIPS): the formation of dense clusters of particles surrounded by a dilute vapor \cite{Cates2015}. The non-equilibrium driving and the many-body repulsive interactions make understanding the insertability of active matter difficult.

It has been shown that some frameworks and ideas from equilibrium statistical mechanics, \emph{i.e.} of passive matter, can be leveraged to understand the thermodynamics of active matter \cite{Fodor2016}. Motivated by these successes, we here begin to investigate an unexplored avenue: the relationship between thermodynamics and insertion space (IS), or, in general, statistical geometry, that was established for equilibrium hard-spheres \cite{Speedy1980,Speedy1981,Speedy1982,Speedy1991A,Speedy1991B,Sastry1998,Debenedetti1999,Bowles2000}. Remarkably, there exist \textit{exact} relations, largely attributed to R. J. Speedy, connecting both the equilibrium compressibility $Z = \beta P V/N$ and, separately, the total chemical potential ${\mu}$ to the averaged insertion space, \emph{i.e.,} the room available to place another sphere averaged over all configurations \cite{Speedy1980,Speedy1981} (Fig. \ref{fig:ISExample}):
\begin{align}
        &Z_G := 1 +  \frac{\sigma \langle S_I \rangle}{2d \langle V_I \rangle}, \qquad {\mu}_G := -\ln \frac{\langle V_I \rangle}{V},
        \label{eq:Speedyrelations}\\
        &\underset{\text{At equilibrium}}{\underbrace{{Z_G = Z \equiv \beta P V/N, \qquad  \mu_G = \mu_\text{exc}= \beta \mu - \ln\frac{N\Lambda^d}{V} }}}
        \label{eq:SpeedyrelationsEqui}
\end{align}
where $\beta = (k_B T)^{-1}$ is the inverse temperature, $P$ is the bulk pressure, $V$ is the total volume of the space, $N$ is the number of constituents, $\sigma$ is the particle diameter (or repulsion range), $d$ is the spatial dimension, $\Lambda$ is the thermal de Broglie wavelength, $\langle V_I \rangle$ is the averaged (over all values in all configurations) total insertion volume, and $\langle S_I \rangle$ its averaged concomitant surface. The insertion volume is expressed in terms of the averaged (over all values across all configurations) number of cavities and cavity size as $\langle V_I \rangle = \langle N_\Delta \rangle \langle \Delta \rangle$ \cite{Speedy1991A,Sastry1998}. 

The appeal of the above relations, is that--at equilibrium--each of the LHS of \eqref{eq:Speedyrelations} have clear thermodynamic interpretations \eqref{eq:SpeedyrelationsEqui} while the RHS of \eqref{eq:Speedyrelations} are purely geometric. However, it is not straightforward how one should view the above relations in light of active matter: systems far from equilibrium. Already, it has been shown that an equation of state, $Z$, with the equilibrium interpretation, is the exception and not the rule for active systems \cite{Solon2015}. Thus, if \eqref{eq:SpeedyrelationsEqui} were to somehow hold the interpretation of the thermodynamic objects must change. Though, before this, a systematic investigation into the geometric objects in \eqref{eq:Speedyrelations} for active systems needs to be done. An important anchoring point is that the determination and interpretation of the IS quantities are purely geometric, and remain concrete far away from equilibrium. Here, we determine the IS and its statistics for repulsive active matter models in low dimensions.

\begin{figure*}[ht!]
    \centering
    \includegraphics[width=\linewidth]{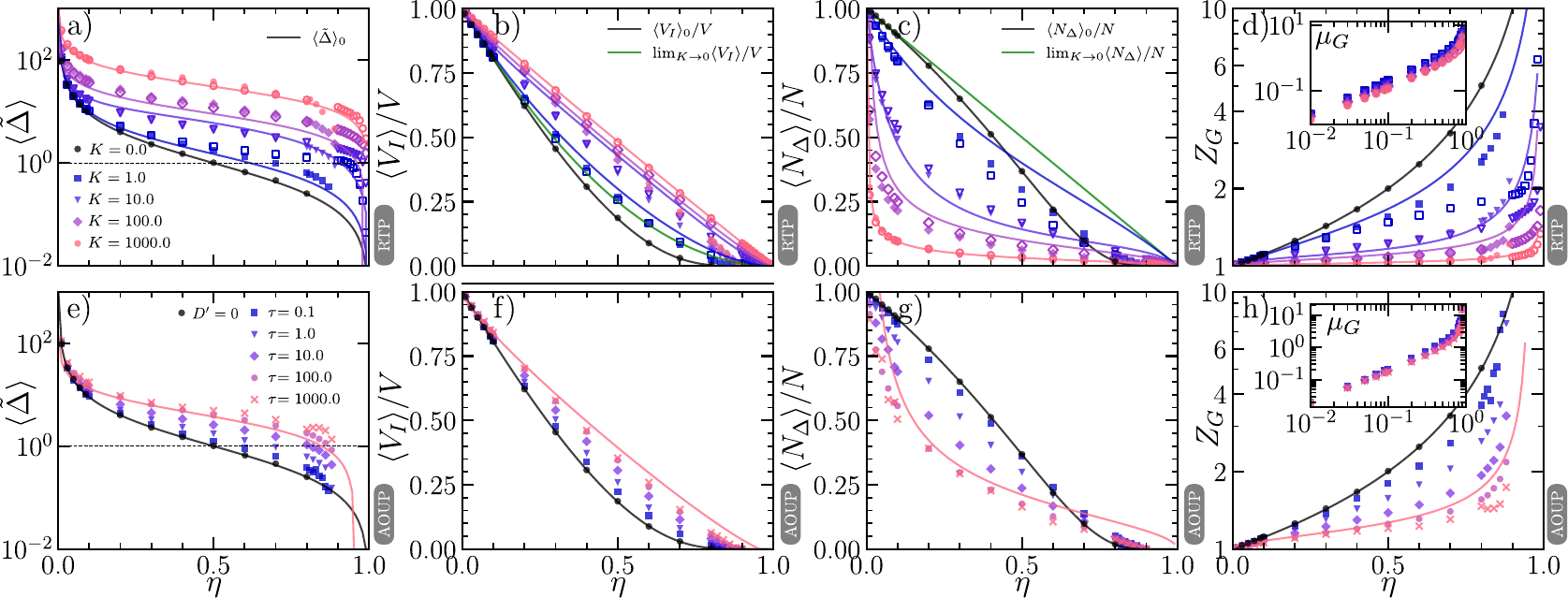}
    \caption{Insertion space (IS) for repulsive active particles in $d=1$. a) Averaged normalized cavity size as a function of packing fraction $\eta$, as determined from off-lattice (filled points) and on-lattice (unfilled points) simulations of RTPs at various $K$. Analytical results are solid lines; for $K\leq 10.0$ the lines are the weak activity mean-field expressions while for $K > 10.0$ we show the high-activity expressions (\ref{eq:activeDELTA}-\ref{eq:activeVI}). Exact passive expressions are from \eqref{eq:ISExact1D}, and are shown along with strict passive simulations ($K=0$). b) The averaged total insertion volume divided by the system volume. c) Averaged number of insertion cavities per particle as a function of $\eta$. d) The geometric compressibility, $Z_G$, as a function of $\eta$. Inset is the geometric chemical potential $\mu_G$ against $\eta$. e) - h) The same quantities as for a)-d) but for the AOUP (off-lattice) model. High-activity analytics (for $\tau = 1000$) are (\ref{eq:activeDELTA}-\ref{eq:activeVI}) with $K \simeq \sqrt{D'\tau}/\sigma$. Parameters (RTPs): $N=1024$, $\mu = 1$ (off-lattice), $\epsilon = 10$ (off-lattice), and $\sigma = 1$; AOUPs: $N=1024$, $\mu =1$, $\epsilon = 10$, $D' = \frac{1}{2}$ and $\sigma =1$.}
    \label{fig:1DRTPs}
\end{figure*}

Firstly, in $d=1$, the averaged total number of cavities $\langle N_\Delta \rangle$ can be written in terms of $P_C$, the contact probability of two particles, as 
\begin{equation}
    \langle N_\Delta \rangle = N(1 -P_C),
    \label{eq:NDELTA}
\end{equation}
 with $\langle N_\Delta \rangle = N$ in the dilute limit (packing fraction $\eta = N \sigma/V \rightarrow 0$) and $\langle N_\Delta \rangle = 0$ at full packing ($\eta \rightarrow 1$). The averaged total insertion volume is, as a function of $P_C$,
 \begin{equation}
     \langle V_I \rangle = V + N \sigma(P_C - 2),
     \label{eq:VI}
 \end{equation}
with $\langle V_I \rangle = V - N (2\sigma)$ in the dilute case, where $2 \sigma$ is the volume excluded by one particle, and $\langle V_I \rangle = 0$ at full packing. Then, based on simple manipulations of (\ref{eq:NDELTA},\ref{eq:VI}), the dimensionless averaged cavity size $\left\langle \Delta \right\rangle/\sigma \equiv \langle \Tilde{\Delta} \rangle$ is given as:
\begin{equation}
\begin{aligned} \label{eq:LatticeDelta}
        \langle \Tilde{\Delta} \rangle &= \frac{\langle V_I \rangle}{\sigma \langle N_\Delta\rangle} &= \frac{1+\eta(P_C - 2)}{\eta(1-P_C)}.
\end{aligned}
\end{equation}
The above exact relations, expressing the insertion observables in terms of $P_C$, appear to be new and are consistent with the expressions in \cite{Speedy1981}.

In one dimension, $\langle S_I \rangle = 2 \langle N_\Delta \rangle$ exactly, implying that the geometric compressibility $Z_G$ \eqref{eq:Speedyrelations} can be simplified as $
Z_G {=} 1 + \langle \tilde{\Delta} \rangle^{-1}$\cite{Speedy1981}. The above observables for the equilibrium one-dimensional hard-sphere model, \emph{i.e.,} the Tonks gas \cite{Tonks1936}, are given exactly as \cite{Tonks1936,Speedy1982,Speedy1991A}:
\begin{equation}
\begin{aligned}
    \langle \tilde{\Delta} \rangle_0 = \frac{(1-\eta)}{\eta}, \quad & \langle N_\Delta \rangle_0 = N \exp \left(-\frac{1}{\langle \tilde{\Delta} \rangle_0}\right), \\
    \langle  V_I \rangle_0 = V (&1-\eta) \exp \left(-\frac{1}{\langle \tilde{\Delta} \rangle_0}\right),
    \label{eq:ISExact1D}
\end{aligned}
\end{equation}
with a corollary that $P_C^\text{eq} = 1 - \exp(- \langle \tilde{\Delta} \rangle_0^{-1})$.

To make analytical progress for the active model we consider repulsive run-and-tumble particles (RTPs) in $d=1$. The model consists of $N$ RTPs, of diameter $\sigma$, on a lattice of $n = V/\sigma \in \mathbb{Z}^+$ sites. In this model the run lengths $k_i$, for particle $i$, are exponentially distributed with rate-parameter $K^{-1} \geq 0$ and during the runs the particles have a constant self-propulsion speed $v_0$ \cite{Mori2020} (see Supplemental Material \cite{supp} for all model details). We note that $K \sim l_p/\sigma$ is to be interpreted as some persistence length divided by the particle size. At equilibrium ($K\rightarrow0$) the model reduces to the simple symmetric exclusion process (SSEP) \cite{Derrida2011}. The RTP particle model can be mapped to a mass transfer model with fluctuating directed bonds (as is done in the Dandekar-Chakraborti-Rajesh (DCR) model \cite{Dandekar2020}), which we exploit here.
The neighboring probability $P_C$ can be broken down as $4 P_C= {p_{\rightarrow,\leftarrow}+2p_{\rightarrow,\rightarrow}+p_{\leftarrow,\rightarrow}},$
where $p_{\rightarrow,\leftarrow}$ is the probability that two particles are next to each other with the left one moving right and the right one moving left, $p_{\rightarrow,\rightarrow}\equiv p_{\leftarrow,\leftarrow}$ is the probability that two particles are next to each other where both particles move in the same direction, and $p_{\leftarrow,\rightarrow}$ is the probability to be next to each other with the left particle moving left and the right particle moving right. The expressions for $p_{\rightarrow,\leftarrow}$, $p_{\rightarrow,\rightarrow}$ and $p_{\leftarrow,\rightarrow}$ are difficult to obtain exactly, and so we resort to a mean-field approximation (as is done in \cite{Dandekar2020}, see Supplemental Material for details). For weak activity, \emph{i.e.,} $K \lesssim 1$, the mean-field $P_C$ reads (up to $O(K^3)$) as, 
\begin{equation}
\begin{aligned}
       P_C&\underset{K\lesssim1}{=} \eta+\frac{\eta(\eta-1)^2K}{2} +\frac{(\eta-1)^2\eta(4\eta^2-7\eta+1)K^2}{8}.
       \label{eqn:WeakP0}
\end{aligned}
\end{equation}

For the high-activity (large persistence) regime naively expanding $P_C$ for small $K^{-1}<<1$ is pathological (as noted in \cite{Dandekar2020}). The main reason for this is that the mean-field treatment assumes bonds (orientations between close particles) are independent and thus cannot--by design--represent the long, strongly correlated domains of like orientation that arise at large persistence. A better high-activity starting point is the coalescence-fragmentation picture in the DCR model (section IV of \cite{Dandekar2020}) which focuses on the interfaces of co-moving particle clusters. In this picture, in the steady-state, there is a balance in the rates of clusters joining and breaking resulting in the density of interfacial bonds (jammed interfaces) going as $\zeta^{\ast} = \sqrt{K^{-1}\langle \tilde{\Delta} \rangle_0}$ and the typical unoccupied cavity size going as $K \zeta^\ast$. Then the contact probability is simply
$P_C \underset{K>>1}{=} 1 - \zeta^\ast$. In the full packing regime both the weak and high activity limits obey $P_C(\eta \rightarrow 1) = 1$, yet in the dilute regime $P_C(\eta \rightarrow 0) \sim 0$ for $K\rightarrow0$ while $P_C(\eta \rightarrow 0) \sim \mathcal{O}(1)$ for $K\rightarrow \infty$ (high persistence limit taken first). 

The dimensionless averaged cavity size \eqref{eq:LatticeDelta} in the presence of weak and strong activity is then, respectively,
\begin{equation}
\begin{aligned}
       \langle \tilde{\Delta}\rangle&\underset{K\lesssim1}{=}\langle \tilde{\Delta} \rangle_0+\frac{K}{2}(1-\eta) -\frac{K^2(1-\eta)(1-5\eta+2\eta^2)}{8}, \\
       \langle \tilde{\Delta}\rangle&\underset{K>>1}{=} K \zeta^\ast -1 \equiv \sqrt{K \langle \tilde{\Delta} \rangle_0}-1,
       \label{eq:activeDELTA}
\end{aligned}
\end{equation}
with $\langle N_\Delta \rangle$ and $\langle V_I \rangle$, also given, respectively, as:
\begin{equation}
    \begin{aligned}
        \frac{\langle N_\Delta \rangle}{N} &\underset{K\lesssim1}{=} (1 - \eta) - \frac{K\eta (1-\eta)^2}{2} \left( 1  - \frac{K(1-7\eta + 4\eta^2)}{4}\right), \\
        \frac{\langle N_\Delta \rangle}{N} &\underset{K>>1}{=} \zeta^\ast \equiv \sqrt{K^{-1}\langle \tilde{\Delta} \rangle_0},
        \label{eq:activeNDELTA}
    \end{aligned}
\end{equation}
and
\begin{equation}
    \begin{aligned}
       \frac{ \langle V_I \rangle}{V} &\underset{K\lesssim1}{=} (1-\eta)^2 \bigg( 1 + \frac{K\eta}{2} + \frac{K^2 \eta(1-7\eta + 4\eta^2)}{8} \bigg), \\
       \frac{ \langle V_I \rangle}{V} &\underset{K>>1}{=} 1- \eta(1+\zeta^\ast) \equiv 1 - \eta\left(1+\sqrt{K^{-1}\langle \tilde{\Delta} \rangle_0} \right).
        \label{eq:activeVI}
    \end{aligned}
\end{equation}
 These relations have two strong physical implications: (i) the relations \eqref{eq:activeDELTA} show that--to leading order--the averaged cavity size is always higher at higher activity, (ii) relations \eqref{eq:activeNDELTA} imply that the presence of activity tends to reduce, relative to the passive case, the total number of insertion cavities, and (iii) relations \eqref{eq:activeVI} imply that the presence of activity increases, relative to equilibrium, the averaged total insertion space across the entire packing fraction. Another way to think of this is through the increased probability of insertion $\mathcal{P}_I \equiv \langle V_I (K,\eta) \rangle/V$ \cite{Speedy1981}. Moreover, in the limit of extreme activity ($K\rightarrow \infty$), we obtain a rather simple relationship $\langle V_I \rangle/V =1 - \eta$ which is even simpler than both the exact and mean-field-limit equilibrium expressions. Due to $\langle \tilde{\Delta} \rangle > \langle \tilde{\Delta} \rangle_0$, for $K>0$, we predict a lower $Z_G$ across the full range of $\eta$. Interestingly, in the limit $\lim_{K\rightarrow\infty} Z_G = 1$ which is equivalent to the compressibility for an equilibrium ideal gas. Additionally, we predict a lower geometric chemical potential $\mu_G$ on the basis that $\langle V_I (K,\eta) \rangle > \langle V_I (\eta)\rangle_0$ $\forall K, \eta$. These insights and the relations (\ref{eq:activeDELTA}-\ref{eq:activeVI}) are our main theoretical results.
To test our theoretical predictions (\ref{eq:activeDELTA}-\ref{eq:activeVI}) we perform on- and off-lattice numerical simulations of repulsive RTPs in $d=1$ (Fig. \ref{fig:1DRTPs}). The implementation of the on-lattice model consists of RTPs hopping between adjacent sites and tumbling at exponentially-sampled rates with rate parameter $K^{-1}$, with a successful move to a new site only if that site is unoccupied. The off-lattice model has essentially the same ingredients, with the exception of continuous space and strict volume-exclusion being modeled by a repulsive (WCA) potential  (see Supplemental Material). In the simulations, for a given configuration $\omega$ the IS can be quantified through the insertion profile $\rho_I(x; \omega) := 1- \Theta\left(\sum\limits_{i=1}^N \Theta(\sigma- |x-x_i|) \right)$, where $\Theta(\ldots)$ is the Heaviside function. By assuming that particles are ordered on the line as $i=1,\ldots,N$ the insertion cavity $\tilde{\Delta}_{i,i+1}(\omega)$, between two neighboring particles $i$ and $i+1 \mod N$, the instantaneous number of insertion cavities per particle $N_\Delta(\omega)/N$, and the total insertion space $V_I(\omega)/V$ are exact (see Supplemental Material).

Our numerical results are, on the whole, in very good agreement with the theoretical mean-field predictions (Fig. \ref{fig:1DRTPs}). For $\langle \tilde{\Delta} \rangle$ the agreement between the on- and off-lattice simulations is excellent for $\eta \lesssim 0.5$ but worsens for both higher $\eta$ and lower $K$ (Fig. \ref{fig:1DRTPs}(a)). The reason for the former is that at higher packing, the off-lattice simulations are able to better sample smaller (than $\sigma=1$) insertion cavities which is difficult in the on-lattice simulations with a single-site per particle. For the diminishing discrepancy at higher $K$: for a particular $\eta$ the increase in activity increases the mean cavity size, which tends to be higher than $\sigma$ and which the on-lattice simulations are better able to sample, however as the system becomes more dense the average cavity size will naturally start to shrink where, again, the on-lattice simulations will struggle to sample these lower values. As expected, as the activity weakens, with $K \rightarrow 0$, the data approaches the known passive result $\langle \tilde{\Delta} \rangle_0$ \eqref{eq:ISExact1D} (see also leading order term of weak-activity expression \eqref{eq:activeDELTA}). 

Gratifyingly, the numbers for $\langle V_I \rangle/V$ (Fig. \ref{fig:1DRTPs}(b)) at high-activity align excellently with our prediction that $\langle V_I \rangle/V = 1 -\eta$. We note that the passive limit of  \eqref{eq:activeVI} is in much better agreement for the $K=1.0$ data than the full weak-activity expression (up to $\mathcal{O}(K^3)$), this suggests that (at $K=1$) all $K$-terms should approximately cancel to leave $\langle V_I \rangle/V = (1-\eta)^2$. For the averaged number of insertion cavities $\langle N_\Delta \rangle/N$ (Fig. \ref{fig:1DRTPs}(c)) we note the excellent agreement between the simulations and the mean-field prediction, which worsens for lower $K$. Interestingly, at higher activity, \emph{e.g., } $K=1000$, there is a rapid decline of the number of insertion cavities at small $\eta$ which then flattens out. The interpretation of this is that the insertion space tries to be as connected (in the percolation sense) as possible. We find that both the geometric compressibility and geometric chemical potential are lower for higher $K$ (Fig. \ref{fig:1DRTPs}(d)). Lastly, we find that, out-of-equilibrium, care must taken with the averaging, where naive averages of $N_\Delta$ and $\Delta$ are no longer statistically independent (see Supplemental Material).

\begin{figure}[t!]
    \centering
    \includegraphics[width=0.8\linewidth]{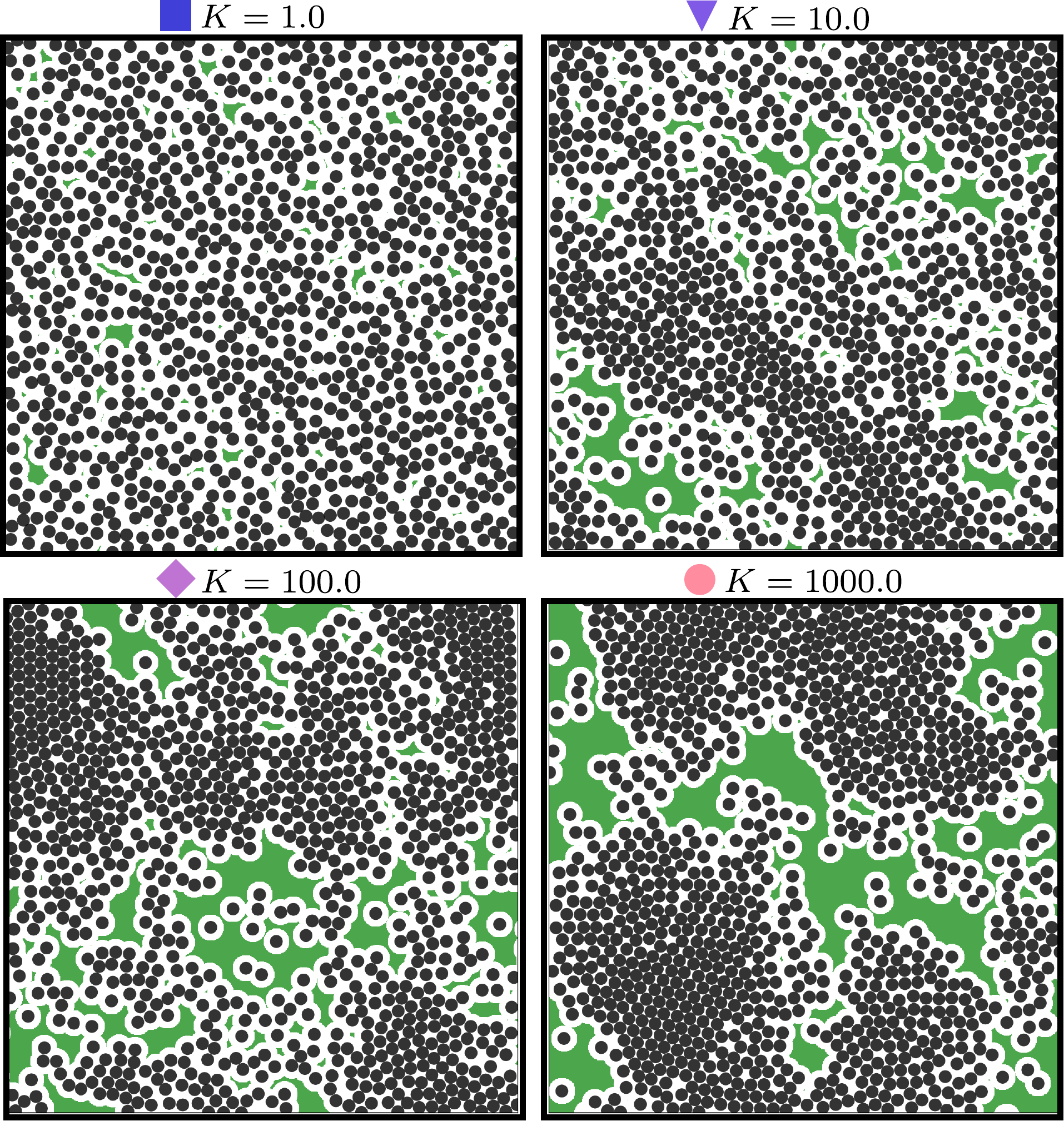}
    \caption{Snapshots from $d=2$ off-lattice simulations of $N=500$ RTPs and insertion cavities for varying $K$ at $\eta=0.4$.}
    \label{fig:2DRTPSnaps}
\end{figure}

We next wondered whether our insights carried over to a different model of repulsive active matter. Thus, we performed off-lattice numerical simulations of repulsive active Ornstein Uhlenbeck particles (AOUPS) in one-dimension. To tune activity we varied the persistence time of the OU process $\tau$, with passivity taken at $\tau \rightarrow 0$. Overall, we find that the insertion space in the AOUP case closely follows that for RTPs: the average insertion cavity and averaged total insertion volume are always higher for increasing activity, whilst the averaged number of cavities, geometric compressibility, and geometric chemical potential tend to decrease with activity (see Fig. \ref{fig:1DRTPs}). For the AOUP model a typical persistence length is $ \ell_p = v_{\mathrm{rms}} \tau $ with $ v_{\mathrm{rms}}=\sqrt{\langle v^2\rangle}=\sqrt{D'/\tau} $ so that $ \ell_p=\sqrt{D'\tau} $ and one can define $ K=\sqrt{D'\tau}/\sigma$. To our surprise, we find that the high activity expressions of the IS observables of $\langle \tilde{\Delta} \rangle$ (\ref{eq:activeDELTA}-\ref{eq:activeVI}) with the replacement $K = \sqrt{D' \tau}/\sigma$, derived for an on-lattice RTP model, quantitatively describes the off-lattice AOUP numerical data for $\tau = 1000$ almost as well as the off-lattice RTP data. 

Thus far, we have only considered one spatial dimension. We now determine how the insertion space changes for repulsive active particles in $d=2$. Analytical results in higher dimensions present significant challenges, even in equilibrium \cite{Speedy1981,Sastry1997}. The mathematical challenge is that, unlike in $d=1$, different particles can form the boundary of a single cavity and thus cavities cannot be labeled by particle indices consistently between different configurations. One certain difference in higher dimensions is that the number of cavities per particle approaches (in the dilute limit) $\langle N_\Delta \rangle \rightarrow 1$, in contrast with $\langle N_\Delta \rangle \rightarrow N$ for $d=1$ \cite{Speedy1991B}. Another fact is that $\langle S_I \rangle/\langle N_\Delta \rangle$, the typical surface area of the insertion cavity, is no longer just a constant.

As there are no analytical relations (on the lattice) in $d=2$ we here only perform off-lattice simulations \footnote{A straightforward way of numerically implementing a two-dimensional repulsive RTP (lattice) model is the four-direction RTP model introduced in \cite{Pietrangeli2024}}. In higher dimensions the determination of the IS is nontrivial. Here we develop a conceptually simple and efficient voxel-based algorithm to determine the IS quantities, which is straightforward to parallelize and also to extend to higher dimensions (see Supplemental Material for the algorithm and benchmark test). In short, our algorithm partitions the space into voxels of side length $l << V^{1/2}$ and removes voxels inside the exclusion zones of particles. If implemented naively (comparing each voxel with every particle) the algorithm is incredibly slow, to avoid this we only check voxels surrounding the particles and leverage the symmetry of the $d-$dimensional cube, greatly improving the computational complexity and attainable resolution. Whilst exact algorithms based on voronoi constructions exist \cite{Sastry1997,Sastry1998,Specht2015}, they are not as easy to implement and to extend to higher dimensions \cite{Dwyer1993}. 

\begin{figure}[t!]
    \centering
    \includegraphics[width=\linewidth]{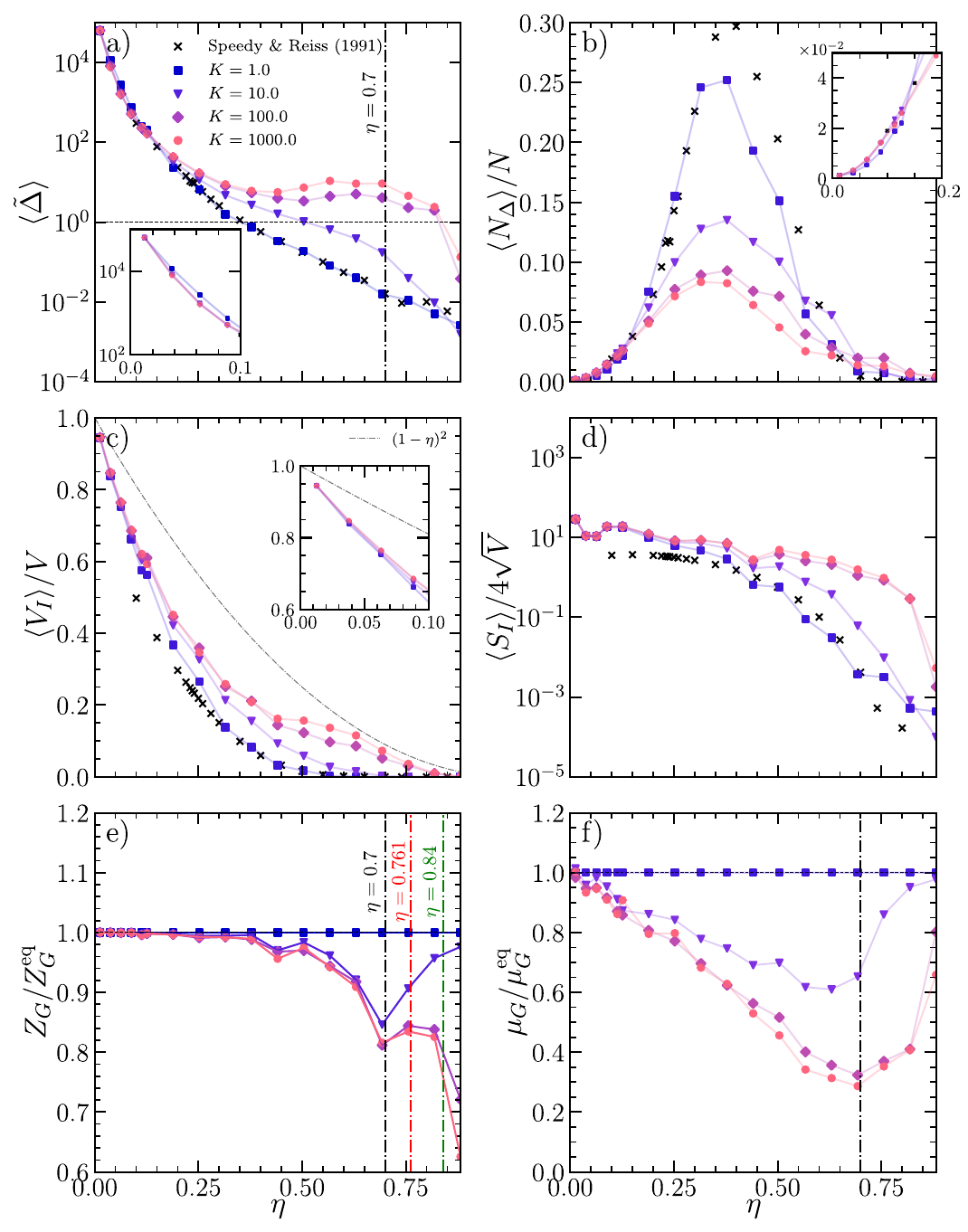}
    \caption{Quantification of the insertion space for off-lattice simulations of $N=1000$ repulsive RTPs in $d=2$. a) Averaged insertion cavity size as a function of $\eta$. Hard disc equilibrium data from \cite{Speedy1991B} are shown as reference. Inset shows a close-up for small packing fractions. b) Number of insertion cavities per particle. c) Averaged total insertion volume. d) The boundary of the insertion volume. e) $Z_G$ \eqref{eq:Speedyrelations} as compared to passivity. The packing fraction $\eta=0.7$ is within 5$\%$ of the freezing transition \cite{Truskett1998} and also the location of liquid-hexatic coexistence at low \textit{Pe} \cite{Digregorio2019}, $\eta=0.761$ is the packing fraction at which the hard-disc fluid freezes \cite{Speedy1991B,Alder1963}, $\eta=0.84$ is the predicted packing fraction at which the metastable fluid to maximally random jammed state transition occurs \cite{Xu2011}. f) The geometric chemical potential compared to passivity.}
    \label{fig:2Ddata}
\end{figure}

Crucially, the off-lattice simulations of RTPS in $d=2$ still show that the total insertion space ($V_I$) remains higher and tends to be more connected as compared with equilibrium (Figs. \ref{fig:2DRTPSnaps} and \ref{fig:2Ddata}). Interestingly, we find that, at high activities $K\geq100$, $\langle \tilde{\Delta} \rangle$ no longer remains strictly monotonic (as in $d=1$) with a wide shelf between $\eta=0.5-0.75$ (Fig. \ref{fig:2Ddata}a)). As expected, our close-to-equilibrium data $K\approx1$ is in excellent agreement with previous IS quantifications in equilibrium hard disc simulations \cite{Speedy1991B}. We observe a tendency for lower averaged number of insertion cavities at higher $K$, with a $\gtrapprox 2$-fold decrease in the maximum at our highest explored activity (Fig. \ref{fig:2Ddata}b)). Interestingly, the location of the maximum of $\langle N_\Delta \rangle$ is approximately the same as in the equilibrium simulations $\eta=0.3-0.4$ \cite{Speedy1991B}. The similar $\eta$ values for the peaks of the number of cavities is likely due to the underlying percolation transition of the hard disc model, where the number of disconnected insertion cavities and particle clusters become equal, though we do not analyze that here \cite{Hoover1979,Speedy1991B}. We note that for small packing fractions, $\eta < 0.1$, the averaged cavity size is slightly smaller for increased activity yet the averaged number of cavities is slightly higher (Figs. \ref{fig:2Ddata}a,b)), which, nonetheless, result in higher total insertion volume, for higher activities, across the entire range of packing fractions (Fig. \ref{fig:2Ddata}c). 

For the boundary of the insertion space $\langle S_I \rangle$ we find it to be a monotonically decreasing function of $\eta$ (due to periodic boundaries) and generally higher for higher activity, though some saturation appears for very high activities $K\geq100$ (Fig. \ref{fig:2Ddata}d)). For the geometric compressibility $Z_G$ we find that the equilibrium $Z^\text{eq}_G$ well approximates the active data for low packing $\eta < 0.4$, but then $Z_G/Z_G^\text{eq}$ starts to rapidly deviate to lower values (Fig. \ref{fig:2Ddata}e)). Interestingly, close to known values of phase transitions (freezing \cite{Truskett1998} and liquid-hexatic coexistence \cite{Digregorio2019}), $\eta \gtrapprox 0.7$, we find that the slope of $Z_G/Z^\text{eq}_G$ abruptly changes sign. Only for the high activity data does it again change sign, at $\eta \approx 0.8$. We also observe a sign change in the slope of the geometric chemical potential $\mu_G/\mu^\text{eq}_G$ (Fig. \ref{fig:2Ddata}f)).

In this letter, we have made the first analytical and numerical investigations into the insertion space of volume-excluding (repulsive) active matter. We show that, across the entire range of packing fractions, the total insertion space is \textit{always} higher in the presence of activity, however weak or strong. This fact is borne out in analytics and simulations for two different well-studied particle models of active matter: RTPs and AOUPs. We expect that our findings to be true for different models of repulsive active matter. We have discovered that activity tends to reduce the total number of insertion cavities, implying that the insertion space remains more connected.
Importantly, our results and predictions regarding the insertion space of active matter are clearly experimentally testable, and could be tested in systems of monodisperse active colloids. Future work, and challenges, include a systematic investigation comparing mechanical definitions of compressibility, pressure, and chemical potential in active systems \cite{Takatori2015,Solon2015,Gaspard2020,Cameron2023,SchiltzRouse2023}. Further, it is not exactly clear how the increased (active) insertion volume should relate to mechanical explanations of MIPS: the slowing down of particles with density results in instabilities that favor clustering \cite{Cates2015}. Other questions are: how is the insertion space related to both the degree of time irreversibility and the entropy production \cite{OByrne2022}? Can the active insertion space reveal the thermodynamic phases of active matter \cite{Klamser2018}? Remaining technical challenges are to find exact, as well as more direct perturbative, relations for the insertion space. We also wonder how the insertion space behaves in the context of other active matter systems, such as non-reciprocal active matter \cite{Klapp2022}, and how it could inform the control of active matter \cite{Davis2024}.

\textit{Acknowledgments:-} We thank Prashant Singh, Mike Cates, and Rob Jack for early discussions. L.K.D. acknowledges the Flora Philip Fellowship at the University of Edinburgh and funding from the Isaac Newton Institute for Mathematical Sciences Postdoctoral Research Fellowship (EPSRC Grant Number EP/V521929/1). L.K.D. thanks the Rosenfeld Foundation for supporting a visit to the NBI, Copenhagen. L.K.D. acknowledges the use of the Edinburgh Compute and Data Facility (ECDF).  

\twocolumngrid

%

\clearpage 
\onecolumngrid 
\pagestyle{empty}
\thispagestyle{empty}

\newcounter{supppage}
\setcounter{supppage}{1}

\whiledo{\value{supppage} < 8}{%
    \clearpage
    \thispagestyle{empty}
    \vspace*{-1in} 
    \hspace*{+\hoffset}\hspace*{+7\oddsidemargin} 
    \noindent\makebox[\paperwidth][c]{
        \vbox to \paperheight{
            \vfill
            \includegraphics[page=\thesupppage, scale=0.99]{supplementalmaterial.pdf}
            \vfill
        }%
    }%
    \stepcounter{supppage}
}

\end{document}